\documentstyle[aps,pre,amsfonts,twocolumn,psfig]{revtex}

\newcommand{\beq}{\begin{equation}}
\newcommand{\eeq}{\end{equation}}
\newcommand{\bea}{\begin{eqnarray}}
\newcommand{\eea}{\end{eqnarray}}

\begin{document}

\title{\bf Geometric approach to chaos in the classical dynamics of
abelian lattice gauge theory}

\author{Lapo Casetti\cite{lapo}}
\address{Istituto Nazionale per la Fisica della
Materia (INFM), Unit\`a di Ricerca del Politecnico di Torino,\\
Dipartimento di Fisica, Politecnico di Torino,
Corso Duca degli Abruzzi 24, I-10129 Torino, Italy}

\author{Raoul Gatto\cite{raoul}}
\address{D\'epartement de Physique Th\'eorique, Universit\'e de Gen\`eve,
24 Quai Ernest-Ansermet, CH-1211 Gen\`eve, Switzerland}

\author{Marco Pettini\cite{marco}}
\address{Osservatorio Astrofisico di Arcetri, Largo Enrico Fermi 5,
I-50125 Firenze, Italy}

\date {\today}
\maketitle

\begin{abstract}
A Riemannian geometrization of dynamics is used to study chaoticity in
the classical Hamiltonian dynamics of a $U(1)$ lattice gauge theory.
This approach
allows one to obtain analytical estimates of the largest Lyapunov exponent
in terms of time averages of geometric quantities. These estimates are compared
with the results of numerical simulations, and turn out to be very close
to the values extrapolated for very large lattice sizes even when the geometric
quantities are computed using small lattices. The scaling of the Lyapunov
exponent $\lambda$ with the energy density $\varepsilon$ is found to be
well described by the law $\lambda \propto \varepsilon^2$.
\end{abstract}

\narrowtext

\section{Introduction}

Classical dynamical aspects of lattice gauge theories have recently attracted
some interest \cite{biro_book,old_works}.
The classical limit of a lattice gauge
theory is interesting both from the point
of view of classical dynamical system theory  and in the perspective of
its quantum counterpart.
As far as the latter aspect is concerned, the interest mainly resides in the
fact that very few non-perturbative tools are available to study quantum gauge
field theories, while in the classical limit it is in principle possible
to exactly simulate the  real time evolution of the system
at any energy. From the point of view of the theory of classical dynamical
systems, lattice gauge theories are highly non-trivial
many-degree-of-freedom Hamiltonian systems which
exhibit a rich and interesting phenomenology. The classical
Hamiltonian dynamics
of such systems is known to be chaotic
\cite{biro_book}; however, a precise characterization of the
different chaotic regimes which may occur in these systems is still
lacking.

Many particular aspects of this general problem have been considered
in the literature, concerning the properties of pure gauge theories and
of theories coupled with matter (mainly Higgs fields), e.g.,
the systematic study of the Lyapunov spectra \cite{biro},
the study of thermalization processes \cite{heinz}, and the relation
between Lyapunov exponents and observable quantities like the
plasmon damping rate \cite{plasmon}.

A particular problem which is still open is the dependence
of the largest Lyapunov exponent $\lambda$ of the pure Yang-Mills lattice
theory on the energy density $\varepsilon$ --- energy per plaquette,
or energy per degree of freedom --- particularly at low $\varepsilon$
\cite{muller,nielsen}.
First, the $\varepsilon$ scaling of $\lambda$
seems different according to the fact that the theory is
abelian --- $U(1)$ --- or non-abelian --- $SU(2)$ and $SU(3)$; in the
latter case two different scalings have
been measured, namely $\lambda \propto \varepsilon^{1/4}$
\cite{lambda1/4} and $\lambda \propto \varepsilon$ \cite{biro_book}, while
in the former case a rapid decrease of $\lambda$ at low $\varepsilon$ was
observed \cite{biro_book}. As we shall see in the following, our results
suggest that in the $U(1)$ case the power law $\lambda \propto \varepsilon^2$ holds.
As pointed out by M\"uller and Trayanov \cite{muller_trayanov}
and subsequently by Nielsen {\em et al.} \cite{nielsen},
such a problem is interesting because
it is tightly related with the problem of the relevance of the chaotic
lattice dynamics for the continuum limit of the gauge
theory. In fact, intrinsic dimensional arguments can be used to show
that the lattice spacing $a$, which can be set equal to one
in all the numerical simulations after a convenient choice of units,
enters the relation between $\lambda$ and $\varepsilon$ as follows:
\beq
a \lambda(a) = f(a \varepsilon(a))~,
\eeq
hence if one observes numerically a power law
$\lambda \propto \varepsilon^k$, the latter can be read as
\beq
\lambda(a) \propto a^{k-1} \varepsilon(a)~.
\eeq
This means that considering a continuum limit $a \to 0$ in which
$\varepsilon (a = 0)$ is finite, corresponding to a finite
temperature of the resulting field, then the Lyapunov exponent
is finite in the limit $a \to 0$ only for the particular
exponent $k=1$. Larger exponents $k > 1$ would imply that
$\lim_{a\to 0}\lambda (a) = 0$, thus leading to a regular continuum theory,
while exponents $k < 1$ would mean that in the continuum theory the
Lyapunov exponent diverges. The linear behavior plays then a very
special role.
As regards non-abelian theories, in Ref. \cite{muller} some evidence is
reported supporting the fact that the correct scaling is the linear one,
$\lambda \propto \varepsilon$,  the other one
($\lambda \propto \varepsilon^{1/4}$) being a spurious result due to
finite-time effects. According to these results the continuum limit,
for small but finite energy densities, of the Lyapunov exponent
of non-abelian lattice gauge theories is finite.
In fact, extracting reliable informations
from numerical simulations in the low-energy regime is very difficult,
mainly because of finite-time effects, which become very important
at small energy densities as the characteristic instability time
scales grow, and of finite-size effects which are known to be
rather large in typical lattice gauge theory simulations.

In the present work we apply a recently proposed formalism
\cite{Pettini,CasettiPettini,prl95,pre96,Rivista},
which is
based on a Riemannian geometrization of Hamiltonian dynamics, to the
classical dynamics of lattice gauge theories. Such a formalism allows one
to relate chaotic dynamics and curvature properties of suitable manifolds,
and to obtain an analytic formula for the Lyapunov exponent in terms
of average curvature fluctuations \cite{prl95,pre96}. The quantities
entering this formula are statistical averages which can be computed
regardless of the knowledge of the dynamics, either by MonteCarlo
or molecular dynamics simulations, or analytically in some cases \cite{pre96}.
As a first step, we apply this formalism to the abelian --- $U(1)$ --- lattice
gauge theory, which is the simplest one,
leaving the non-abelian case to future work.
In the case of a $U(1)$ gauge theory  we perform
a precise numerical measurement of the Lyapunov exponent by simultaneous
integration of the Hamilton's equations and of the tangent dynamics
equations using a precise symplectic algorithm \cite{algo} for
several lattice sizes. In these simulations we measure also the
relevant geometric observables that allow for a characterization of
the dynamics and that enter the above-mentioned
theoretical expression for the Lyapunov exponent. We find that
the analytic estimate compares very well with the outcomes of the numerical
simulations for the Lyapunov exponents. Moreover, we find that the
theoretical estimate is almost free from finite-size effects;
for small energies we find the
law $\lambda \propto \varepsilon^2$ already when inserting in the formula
 the values of the geometric
observables obtained with small ($4\times 4 \times 4$) lattices, while
the numerical values of the Lyapunov exponents are affected by
large finite-size effects.

The paper is organized as follows: in Section \ref{sec_geom} we review
very briefly the geometric theory of Hamiltonian chaotic dynamics;
in Section \ref{sec_model} we describe the model and the observables
studied. Section \ref{sec_results} is devoted to the presentation and the
discussion of the results, and in Section \ref{sec_remarks} we draw
some conclusions and we outline some future developments.

\section{Geometry and chaotic dynamics}
\label{sec_geom}

Let us now recall very briefly the main points about the geometric
theory of Hamiltonian chaos. Details can be found in Refs.
\cite{Pettini,CasettiPettini,prl95,pre96}.
Despite the fact that this theory is still
at its beginning, it has already proved useful not only in its
original context, but also in connection with the problem of the
relationship between dynamics and statistical mechanics (in
particular phase transitions) \cite{cccp,gatto,phi4_2d}.

Hamiltonian dynamics can be rephrased in geometrical terms owing to
the fact that the trajectories of a dynamical system with quadratic
kinetic energy can be seen as geodesics of a suitable Riemannian
manifold. There are several choices for the ambient manifold as well as
for the metric tensor. As already discussed in Ref.
\cite{CasettiPettini,prl95,pre96} a particularly useful ambient space
is the enlarged configuration space-time $M\times {\Bbb R}^2$, i.e.
the configuration space $\{q^1,\ldots,q^i,\ldots,q^N\}$ with two additional
real coordinates $q^0$ and $q^{N+1}$. In the following $q^0$ will be identified
with the time $t$. For standard Hamiltonians ${\cal H}=T+V({\bf q})$ where
$T=\frac{1}{2}a_{ij}\dot q^i\dot q^j$, this manifold,
equipped with Eisenhart's metric $g_E$,
has a semi-Riemannian (Lorentzian) structure ($\det g_E = -1$).
The arc-length is given by
\begin{equation}
ds^2=a_{ij}dq^i  dq^j - 2V({\bf q})(dq^0)^2 + 2dq^0 dq^{N+1}~,
\end{equation}
where both $i$ and $j$ run between $1$ and $N$. Let us restrict to geodesics
whose arc-length parametrization is  affine,
i.e. $ds^2=c_1^2 dt^2$; simple algebra shows that the geodesic equations
\begin{equation}
\frac{d^2q^\mu}{ds^2}+\Gamma^\mu_{\nu\lambda}\frac{dq^\nu}{ds}
\frac{dq^\lambda}{ds}=0 ~~~~~\mu,\nu,\lambda=0\ldots N+1~,
\end{equation}
become Newton equations (without loss of
generality $a_{ij}=\delta_{ij}$ is considered)
\begin{equation}
\frac{d^2q^i}{dt^2}  =  - \frac{\partial V}{\partial q_i}
\label{eqgeoit}
\end{equation}
for $i=1\ldots N$, together with two extra equations for
$q^0$ and $q^{N+1}$ which can be integrated to yield
\begin{mathletters}
\label{excoord}
\begin{eqnarray}
q^0  & = & t \label{q0(t)} \\
q^{N+1} &  = & \frac{c_1}{2} t+ c_2 - \int_0^t L({\bf q},\dot{\bf q})\,dt
\label{qN+1(t)}
\end{eqnarray}
\end{mathletters}
where $L({\bf q},\dot{\bf q})$ is the Lagrangian, and $c_1$, $c_2$ are
real constants. In the following we set $c_1 = 1$ in order that $ds^2 = dt^2$
on the physical geodesics.
As stated by Eisenhart theorem \cite{Eisenhart}, the dynamical
trajectories in configuration space are projections on $M$ of the geodesics of
$(M\times{\Bbb R}^2,g_E)$.

In the geometrical framework, the stability
of the trajectories is mapped on the stability
of the geodesics, hence it can be studied by the
Jacobi equation for geodesic deviation
\begin{equation}
\frac{D^2 J}{ds^2} + R(\dot\gamma, J)\dot\gamma = 0~,
\label{eqJ}
\end{equation}
where $R$ is the Riemann curvature tensor, $\dot\gamma$ is the velocity
vector along the reference geodesic $\gamma(s)$,  $D/ds$ is the covariant
derivative and $J$,
 which measures the deviation between nearby
geodesics, is referred to as the Jacobi field.
The stability --- or instability --- of the dynamics, and thus
deterministic chaos, originates from the curvature properties
of the ambient manifold. In local coordinates, Eq. \ref{eqJ} is
written as
\begin{equation}
\frac{D^2 J^{\mu}}{ds^2} + R^\mu_{\nu\rho\sigma}
\frac{dq^\nu}{ds} J^\rho \frac{dq^\sigma}{ds} = 0~,
\end{equation}
and in the case of Eisenhart metric it simplifies to
\begin{equation}
\frac{d^2 J^i}{dt^2} + \frac{\partial^2 V}{\partial q_i \partial q^j}
J^j=0~, \label{dintangEis}
\end{equation}
which is nothing but the usual
tangent dynamics equation for standard Hamiltonians.
The Lyapunov
exponents are usually computed evaluating the rate of exponential
growth of  $ J $ by means of a numerical integration of
Eq. (\ref{dintangEis}) \cite{BenettinGS}.

In the particular case of {\em constant curvature}
manifolds, Eq. (\ref{eqJ}) becomes very simple \cite{doCarmo}
\begin{equation}
\frac{D^2 J^\mu}{ds^2} + K \, J^\mu = 0~,
\label{eqJconst}
\end{equation}
and has bounded oscillating solutions $J \approx
\cos(\sqrt{K}\, s)$ or
exponentially unstable solutions $J \approx \exp(\sqrt{-K}\,
s)$
according to the sign of the constant sectional curvature
$K$, which is
given by
\begin{equation}
K = \frac{K_R}{N-1} = \frac{{\cal R}}{N(N-1)}~,
\label{Kconst}
\end{equation}
where $K_R = R_{\mu\nu}\dot q^\mu \dot q^\nu$ is the Ricci
curvature
and ${\cal R} = R^\mu_\mu$ is the scalar curvature;
$R_{\mu\nu}$ is
the Ricci tensor.
Manifolds with $K < 0$ are considered in abstract ergodic
theory (see e.g. Ref. \cite{Sinai}).
Krylov \cite{Krylov} originally
proposed that the presence of some negative curvature
could be the mechanism actually at work to make chaos in
physical
systems, but in realistic cases the curvatures are neither
found constant
nor everywhere negative, and the straightforward approach
based on Eq. (\ref{eqJconst}) does not apply. This is the
main reason why
Krylov's ideas remained confined to abstract ergodic theory
with few exceptions.

In spite of these major problems, some approximations on Eq.
(\ref{eqJ})
are possible even in the general case.
The key point is that negative
curvatures are not strictly
necessary to make chaos, and that a subtler
mechanism related to the {\em bumpiness} of the ambient
manifold is  actually at work. Upon an assumption of quasi-isotropy
of the ambient manifold, i.e., that the manifold can be obtained
as a small deformation of a constant-curvature space (see Ref. \cite{pre96}
for details), Eq. (\ref{eqJ}) can be approximated by an effective scalar
equation which reads
\begin{equation}
\frac{d^2\psi}{dt^2} + K(t)\,\psi = 0~,
\label{eqHill(t)}
\end{equation}
where $\psi$ is a generic component of the vector $J$ (in this approximation
all the components are considered equivalent), and $K(t)$ is
a stochastic process which models the curvature along the geodesic curve. Such
a stochastic model is defined by
\begin{equation}
K(t) = \langle k_R \rangle +
\langle \delta^2 k_R \rangle^{1/2}\eta(t)~,
\label{Kstoc}
\end{equation}
where $k_R= K_R/N$, $\langle \cdot \rangle$ stands for an average
taken along a geodesic, which, for systems in thermal equilibrium,
can be substituted with a statistical average taken with respect to a
suitable probability measure (e.g.
the micro-canonical or the canonical measure); $\eta(t)$ is a  stationary
$\delta$-correlated  Gaussian stochastic process with zero mean
and variance equal to one. Using Eisenhart metric,
and for standard Hamiltonians, the non-vanishing components of the
Riemann tensor are $R_{0i0j} = \partial_{q_i} \partial_{q_j} V$,
hence the Ricci curvature has the remarkably simple form
\begin{equation}
k_R = \frac{1}{N} \nabla^2 V~,
\label{k_R}
\end{equation}
where $\nabla^2$ is the Euclidean Laplacian operator.
Equation (\ref{eqHill(t)}) becomes
a stochastic differential equation, i.e. the evolution equation
of a random oscillator \cite{VanKampen}. It is worth noticing
that Eq. (\ref{eqHill(t)}) is no longer dependent on the dynamics,
since the random process depends only on statistical averages.
The estimate of the Lyapunov exponent $\lambda$ is then obtained
through the evolution of the second moments of the solution of
(\ref{eqHill(t)}) as
\begin{equation}
\lambda = \lim_{t \to \infty} \frac{1}{2t} \log
\frac{\psi^2(t)+\dot\psi^2(t)}{\psi^2(0)+\dot\psi^2(0)}~.
\end{equation}
As shown in Ref. \cite{prl95,pre96}, this yields the following expression
for $\lambda$:
\begin{equation}
\lambda\,(k,\sigma_k,\tau) =
\frac{1}{2}\left(\Lambda-\frac{4 k}{3\Lambda}\right)~,
\label{formula}
\end{equation}
where
\begin{mathletters}
\begin{equation}
\Lambda =
\left( \sigma_k^2\tau + \sqrt{\frac{64 k^3}{27}+
\sigma_k^4\tau^2}~\right)^{1/3}~,
\end{equation}
\begin{equation}
\tau=\frac{\pi\sqrt{k}}{2 \sqrt{k(k + \sigma_k)} + \pi\sigma_k} ~;
\end{equation}
\end{mathletters}
in the above expressions $k$ is the average Ricci curvature
$k = \langle k_R \rangle$
and $\sigma_k$ stands for the mean-square fluctuation of the Ricci
curvature, $\sigma_k = \langle \delta^2 k_R \rangle^{1/2}$.

The advantages in using the geometric approach to Hamiltonian chaos are
thus evident. In fact, it is possible to give reliable estimates of
the Lyapunov exponent without actually computing the time evolution
of the system: the estimate (\ref{formula}) of $\lambda$ depends only
on statistical averages which can be either computed analytically in
some cases (for instance in the case of the FPU model \cite{prl95})
or, in general, extracted from a Monte Carlo or a dynamical simulation,
as it is the case of the model studied in the present work.

The behavior of the average geometric observables as the
control parameter (e.g. the energy density or the temperature) is varied
conveys an information which goes beyond the possibility of computing the
Lyapunov exponent. In fact, one can look at the random oscillator equation
(\ref{eqHill(t)}) as an effective Jacobi equation for a geodesic flow
on a surface
$M$ whose Gaussian curvature is given by the random process $K(t)$. As long
as nonlinear coupled oscillators are considered, the average Ricci curvature
is positive, hence $M$ can be regarded as a sphere with a fluctuating radius.
In the limit of vanishing fluctuations, one recovers the bounded evolution
of the Jacobi field associated with integrable dynamics. Chaos suddenly
appears as curvature fluctuations are turned on, nevertheless it
it will be ``weak'' as long as $\sigma_k \ll k$, i.e. as long as
$M$ can be considered as a weakly perturbed sphere. On the contrary
as the size of curvature fluctuations becomes of the same order
of the average curvature, $\sigma_k \simeq k$,
$M$ can no longer resemble a sphere, and the dynamics will no longer ``feel''
the integrable limit. Hence we expect the dynamics to be strongly chaotic.
This is by no means a deep explanation of the
existence of weakly and strongly chaotic regimes in Hamiltonian dynamics.
Nevertheless it shows how the simple geometric concepts which enter
the Riemannian description of Hamiltonian chaos, besides providing
effective computational tools, are also useful in helping one's physical
intuition with images and analogies which would be difficult to find
elsewhere.

\section{Model and dynamical observables}
\label{sec_model}

The dynamical system that we are now considering is the classical
lattice gauge theory based on the abelian gauge group $U(1)$.
The Hamiltonian of such a system can be derived in two ways,
either as the dual (Wegner model) of a lattice planar spin system
\cite{Kogut}, or starting from the Kogut-Susskind Hamiltonian
for a generic gauge group $G$ \cite{Kogut,KogutSusskind} and then
specializing to the group $U(1)$. In the latter case we obtain the
same Hamiltonian as in the former case by choosing the $SO(2)$
real matrix representation of the group $U(1)$.

The lattice Lagrangian,
obtained from the Wilson action by fixing the temporal
gauge (all the temporal components of the gauge fields are set to zero)
and then by taking the continuum limit in the temporal direction, is
\beq
{\cal L} = \frac{ag^2}{2} \sum_{{\rm links}} \langle \dot U_{x,\mu},
\dot U_{x,\mu} \rangle
- \frac{1}{ag^2} \sum_{{\rm plaquettes}} \left( 1 - \frac{1}{2} {\rm Tr~}
U_{\mu\nu} \right) ~,
\label{lagrangian}
\eeq
where $U_{x,\mu} \in G$ is a group element defined on a link of a
$d$-dimensional cubic lattice, labeled by the site index $x$ and the
oriented lattice direction $\mu$,
$g^2$ is a coupling constant, $a$ is the lattice spacing, $\langle
\cdot,\cdot \rangle$ stands for the scalar product between group
elements, defined as
\beq
\langle A, B \rangle = \frac{1}{2} {\rm Tr~} (AB^\dagger)~,
\eeq
and $U_{\mu\nu}$ is a shorthand notation for the plaquette operator,
\beq
U_{\mu\nu} = U_{x,\mu} U_{x+\mu,\nu} U_{x+\mu+\nu,-\mu} U_{x+\nu,-\nu}~.
\eeq
We can pass to a standard Hamiltonian formulation by putting
\beq
P = \frac{\partial {\cal L}}{\partial \dot U} = ag^2 \dot U~,
\eeq
thus obtaining
\beq
ag^2{\cal H} = \frac{1}{2} \sum_{{\rm links}} \langle P_{x,\mu},
P_{x,\mu} \rangle
+ \sum_{{\rm plaquettes}} \left( 1 - \frac{1}{2} {\rm Tr~}
U_{\mu\nu} \right) ~.
\label{hamiltonian}
\eeq
The parameters $a$ and $g^2$ can be scaled out, so we set $g = a = 1$.

The Hamiltonian (\ref{hamiltonian}) is the classical Hamiltonian for
a lattice gauge theory with a generic gauge group $G$. Let us now
specialize to the abelian group $G = U(1)$. Choosing the representation
\beq
U = \left(
\begin{array}{cc}
\cos \varphi & \sin \varphi \\
-\sin \varphi & cos \varphi
\end{array}
\right)
\eeq
we have
\beq
P = \dot U = \dot \varphi \left(
\begin{array}{cc}
-\sin \varphi & \cos \varphi \\
-\cos \varphi & -\sin \varphi
\end{array}
\right)
\eeq
and we find
\beq
\frac{1}{2} \langle P, P \rangle = \frac{1}{2} \dot \varphi^2~.
\eeq
To write the plaquette operator, we use the fact that the group
$U(1) \simeq SO(2)$ is abelian. Then the product of the four rotations
is a rotation of the sum of the oriented angles
along the plaquette (the discrete curl of the field $\varphi$)
\beq
\varphi_{x,\mu\nu} = \varphi_{x,\mu} + \varphi_{x+\mu,\nu} -
\varphi_{x+\nu,\mu} - \varphi_{x,\nu}~,
\eeq
and the magnetic energy, i.e., the potential energy of the dynamical
system can be written as
\beq
V = \sum_{{\rm plaquettes}} \left( 1 - \frac{1}{2} {\rm Tr~}
U_{\mu\nu} \right) = \sum_{{\rm plaquettes}} \left( 1 - \cos \varphi_{x,\mu\nu}
\right) ~.
\label{V}
\eeq
Our canonical variables are then the angles $\varphi_{x,\mu}$ and the
angular momenta $\pi_{x,\mu} = \dot \varphi_{x,\mu}$, and the
Hamiltonian (\ref{hamiltonian}) becomes
\beq
{\cal H} = \frac{1}{2} \sum_{{\rm links}} \pi^2_{x,\mu}
+ \sum_{{\rm plaquettes}} \left( 1 - \cos \varphi_{x,\mu\nu} \right) ~.
\label{ham}
\eeq
Constant-energy (microcanonical) simulations have been performed
on three-dimensional lattices --- with lattice sizes ranging
from $4^3$ to $15^3$ --- by integrating
the canonical equations of motion
\begin{mathletters}
\label{eqham}
\bea
\dot \varphi_{x,\mu} & = & \pi_{x,\mu} ~; \\
\dot \pi_{x,\mu} & = & - \frac{\partial V}{\partial \varphi_{x,\mu} } ~,
\label{f=ma}
\eea
\end{mathletters}
where $x$ runs over the $L^3$ lattice sites and $\mu = 1,2,3$,
by using a precise third-order bilateral symplectic
algorithm \cite{algo}.
We remind that symplectic algorithms are integration schemes which exactly
preserve the canonical form of the equations of motion. The
exact Hamilton equations are replaced, in the time discretization procedure,
by a map that is symplectic, hence the discrete-time flow that approximates
the true Hamiltonian flow generated by Hamilton' s equations is still
Hamiltonian. All the geometric constraints on the phase space trajectories
which are enforced by the canonical form of the equations of motion are thus
exactly preserved during the numerical integration procedure and the
Hamiltonian flow is faithfully represented. As a by-product
of this features, symplectic algorithms conserve very well the total energy of
the system: in our simulations relative energy fluctuations were of the
order of $10^{-7} \div 10^{-8}$.
In Eqs. (\ref{eqham}) $V$ is given by Eq. (\ref{V}),
whose explicit expression on a three-dimensional lattice is
\beq
V = \sum_x \sum_{(\mu\nu)}
\left( 1 - \cos \varphi_{x,\mu\nu}\right)~,
\eeq
where
$(\mu\nu) = 12, 13, 23$. The forces (rhs of Eq. \ref{f=ma}) are given by
\beq
- \frac{\partial V}{\partial \varphi_{x,\mu} } = \sum_{\delta = 1,2}
\sin \varphi_{x - \mu - \delta, \mu\mu+\delta} - \sin
\varphi_{x,\mu\mu + \delta} ~.
\label{forces}
\eeq
In order to compute the largest Lyapunov exponent $\lambda$ by the standard
method \cite{BenettinGS}, the tangent dynamics equations (\ref{dintangEis}),
which now reads as
\beq
\ddot J_{x,\mu}+ \sum_y  \sum_\nu
\frac{\partial^2 V}
{\partial \varphi_{x,\mu}   \partial \varphi_{y,\nu} }
J_{y,\nu}=0~,
\label{eqdintang_gauge}
\eeq
have been integrated simultaneously with the Hamilton's equations (\ref{eqham})
by means of the same algorithm. The largest Lyapuonov exponent has then been
computed according to the definition
\beq
\lambda = \lim_{t \to \infty} \frac{1}{t} \log
\frac{\left[|\dot J|^2 (t) + |J|^2 (t)\right]^{1/2}}
{\left[|\dot J|^2 (0) + |J|^2 (0)\right]^{1/2}}~.
\eeq
Here $|J|^2 = \sum_x \sum_\mu
 J^2_{x,\mu}$ is the squared Euclidean norm of the
tangent vector $J$.

According to the discussion of Sec. \ref{sec_geom}, the relevant geometric
observable which is able to characterize the chaotic dynamics of the model
is the Ricci curvature $k_R$,
computed with the Eisenhart metric, defined by Eq. (\ref{k_R}), which now
can be
rewritten as
\beq
k_R = \frac{1}{L^3} \sum_x \sum_\mu \sum_{\nu\not=\mu} [ \cos
\varphi_{x,\mu\nu}
+ cos \varphi_{x-\nu,\mu\nu}~ ].
\label{k_R_gauge}
\eeq
The average and the rms fluctuations of the Ricci curvature $k_R$ have been
computed in all the simulations. The results are presented and discussed
in the next Section.

\section{Results and discussion}
\label{sec_results}

\begin{figure}
\centerline{\psfig{file=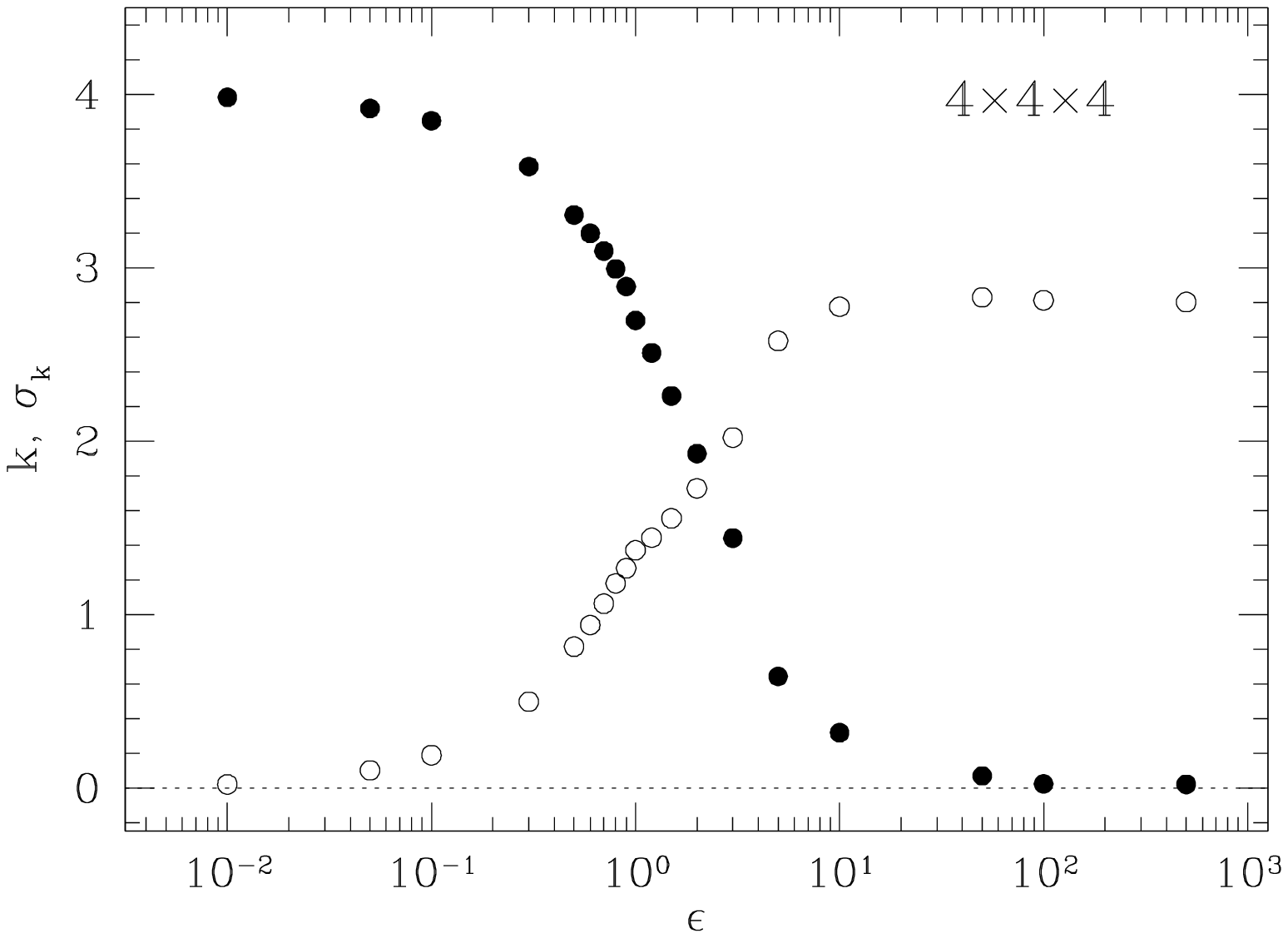,height=6.5cm}}
\centerline{\psfig{file=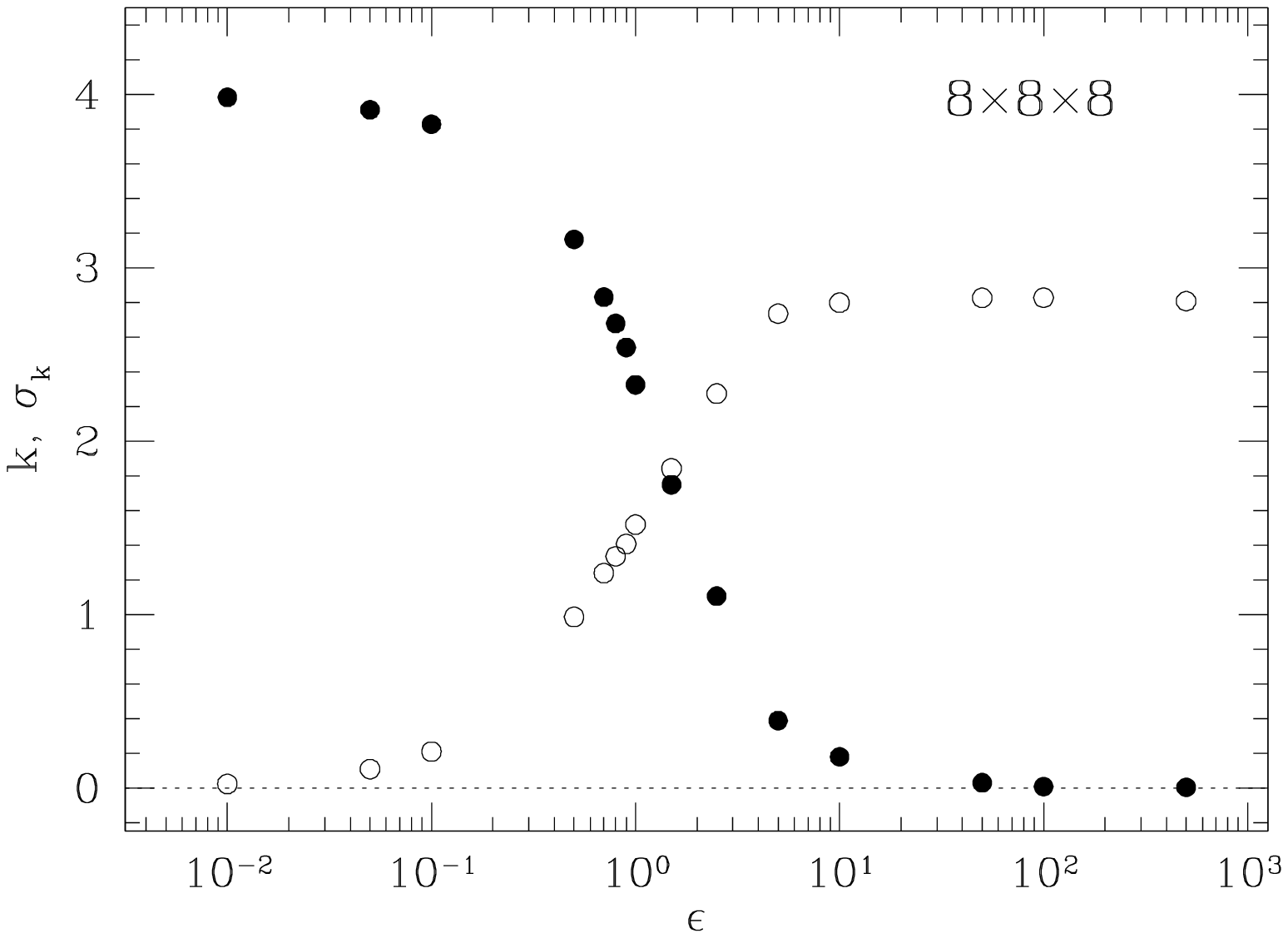,height=6.5cm}}
\centerline{\psfig{file=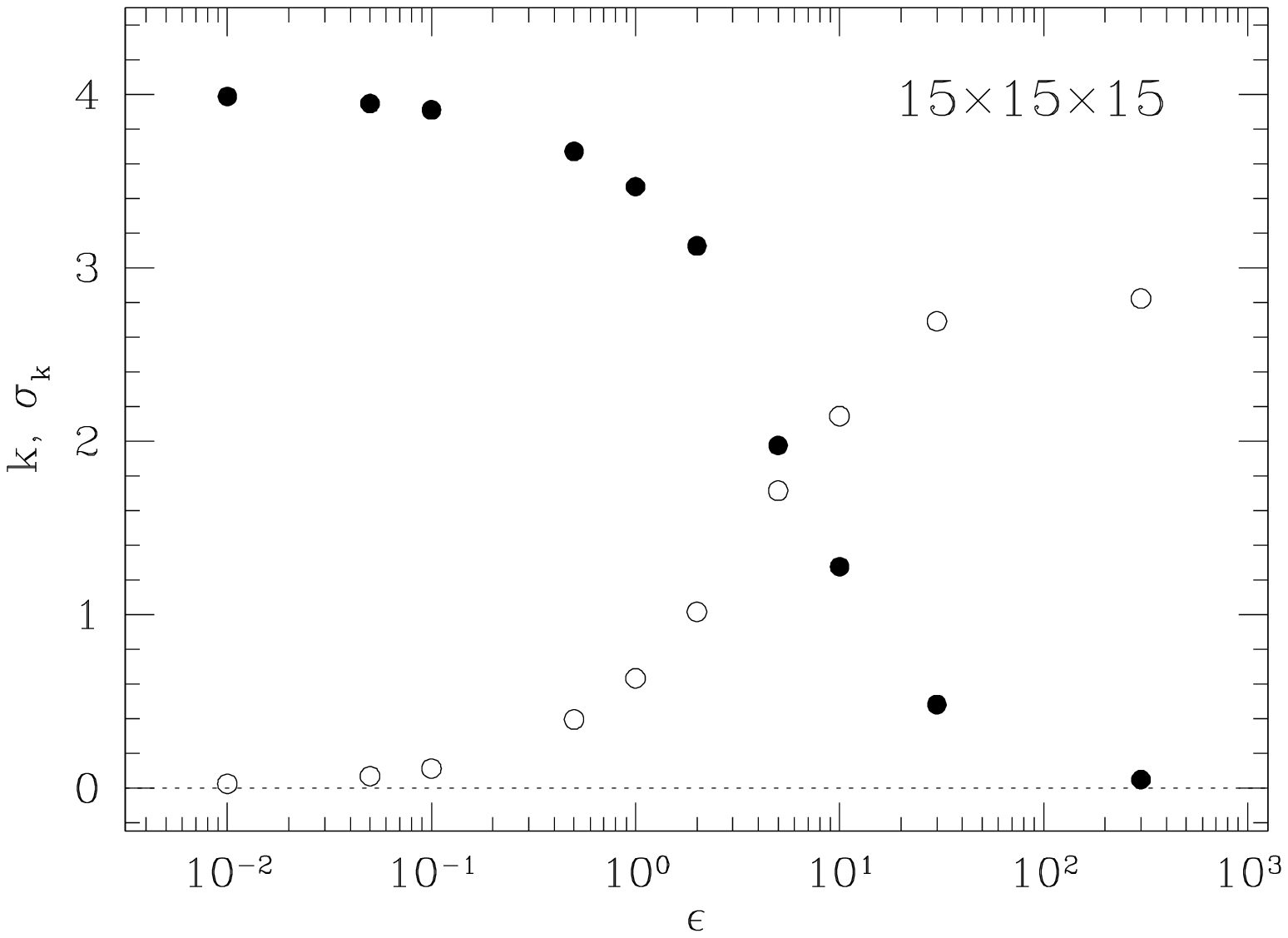,height=6.5cm}}
\caption{Plots of the average $k$ (full circles)
and of the rms fluctuation $\sigma_k$ (open circles) of
the Ricci curvature {\em vs.}
the energy density $\varepsilon$ for three different lattice sizes:
from top to bottom,
$L^3 = 4^3,8^3,15^3$.  Errorbars are smaller than the size of the data points.
\label{fig_k_sigma}}
\end{figure}

The average and the fluctuations of the Ricci curvature (\ref{k_R_gauge})
are plotted in Fig. \ref{fig_k_sigma} against the energy density
$\varepsilon$ for three different lattice sizes.
The resulting patterns reveal that in the low-energy regime the average $k$
is much larger than the fluctuation $\sigma_k$, whereas in the high-energy
regime the converse is true. These results are qualitatively the same as
those that have been recently obtained for a planar (XY) spin model
on a $2-d$ cubic lattice \cite{cccp}. Such a fact is not surprising, since
the present model, possessing a local gauge invariance
under proper planar rotations,
is expected to behave on a $d$-dimensional lattice ---
from a statistical-mechanical point of view --- as
its dual model (invariant under a global symmetry) on a $d-1$-dimensional
lattice  \cite{Kogut}. Moreover, this means that our results are consistent
with the fact that the system undergoes a Bere\v{z}inskij-Kosterlitz-Thouless
(BKT) transition at a finite energy density (of order $\varepsilon \simeq 1$
in our units). However, the analysis of the
geometric quantities is not expected to give sharp indications of the
presence of a transition in the case of a BKT transition. On the contrary
in the case of second-order transitions the fluctuations of the Ricci curvature
exhibit peculiar cusp-like behaviours \cite{cccp,gatto} which are completely
absent here.

\begin{figure}
\centerline{\psfig{file=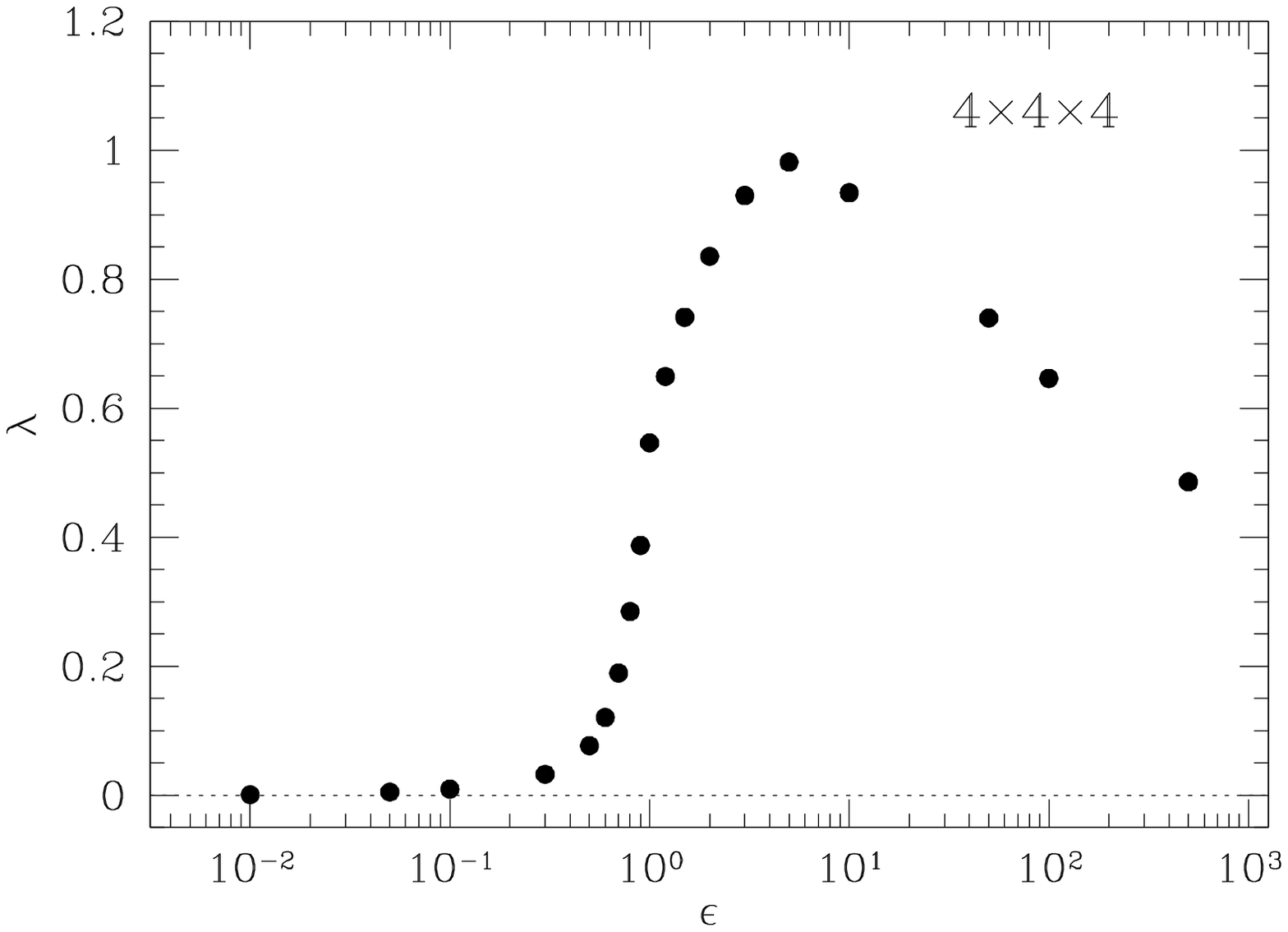,height=6.5cm}}
\centerline{\psfig{file=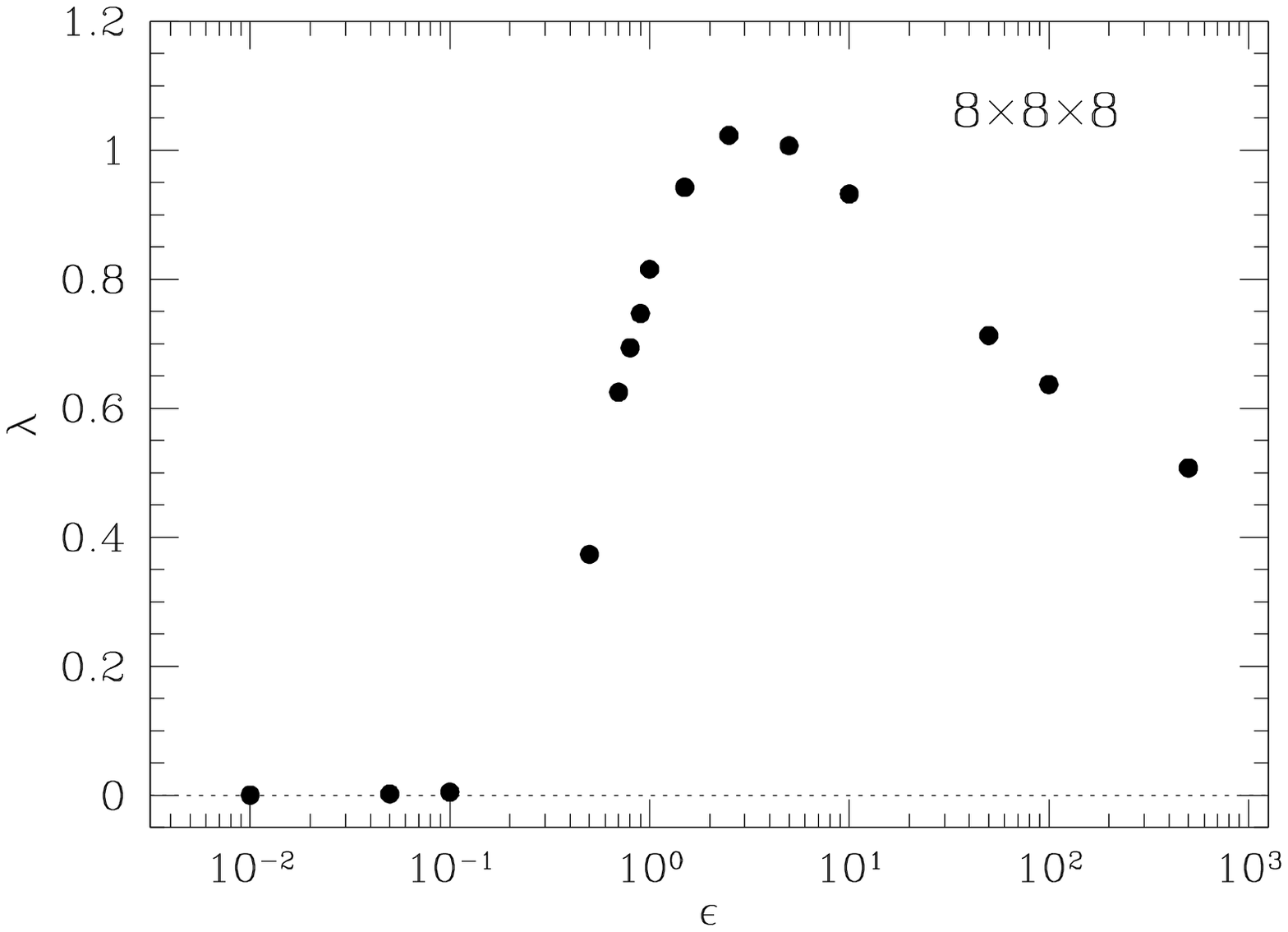,height=6.5cm}}
\centerline{\psfig{file=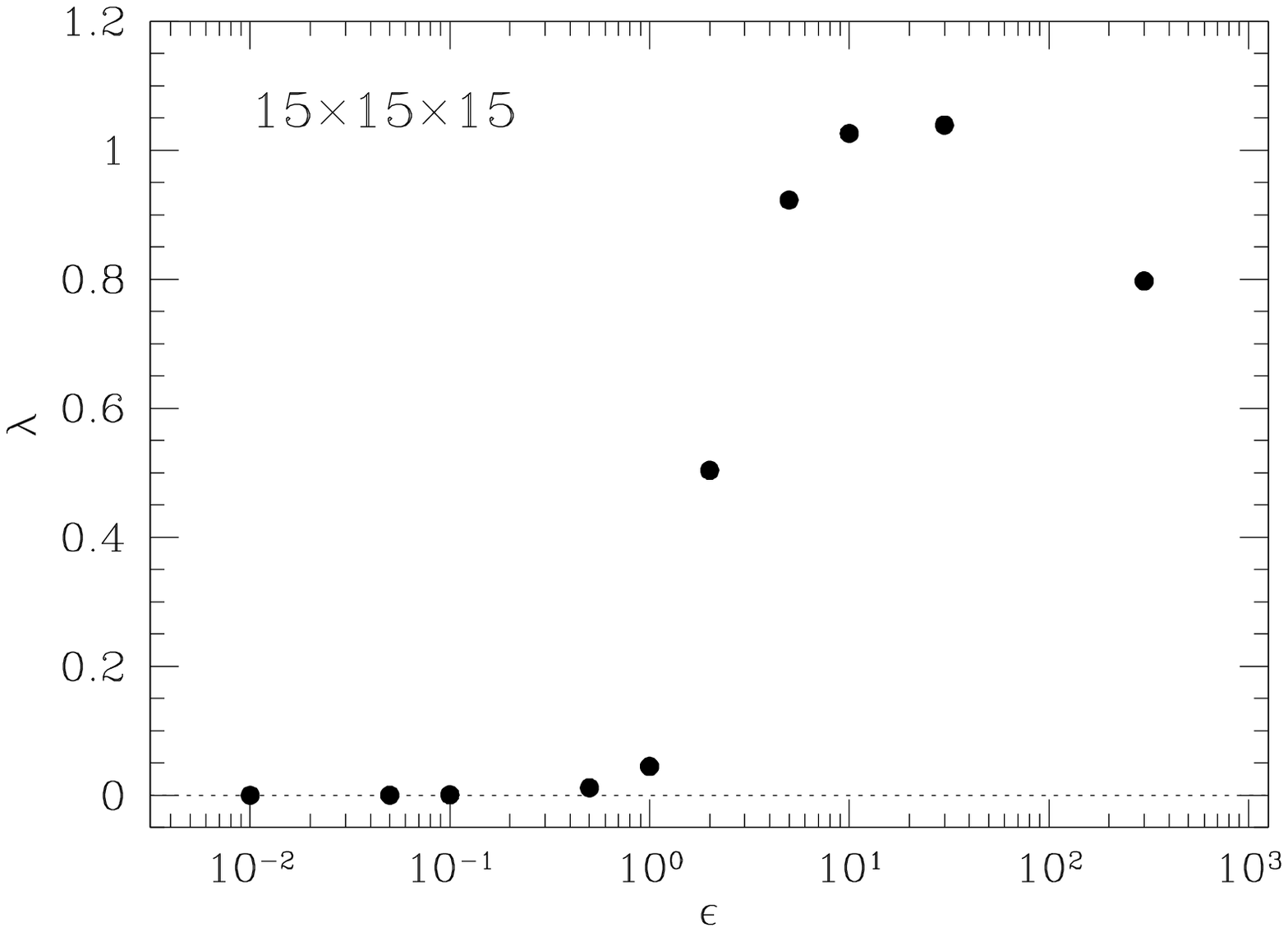,height=6.5cm}}
\caption{Numerically computed largest Lyapunov exponent $\lambda$
{\em vs.} the energy density $\varepsilon$
for three different lattice sizes: from top to bottom,
$L^3 = 4^3,8^3,15^3$.  Errorbars are smaller than the size of the data points.
\label{fig_lyap}}
\end{figure}

\begin{figure}
\centerline{\psfig{file=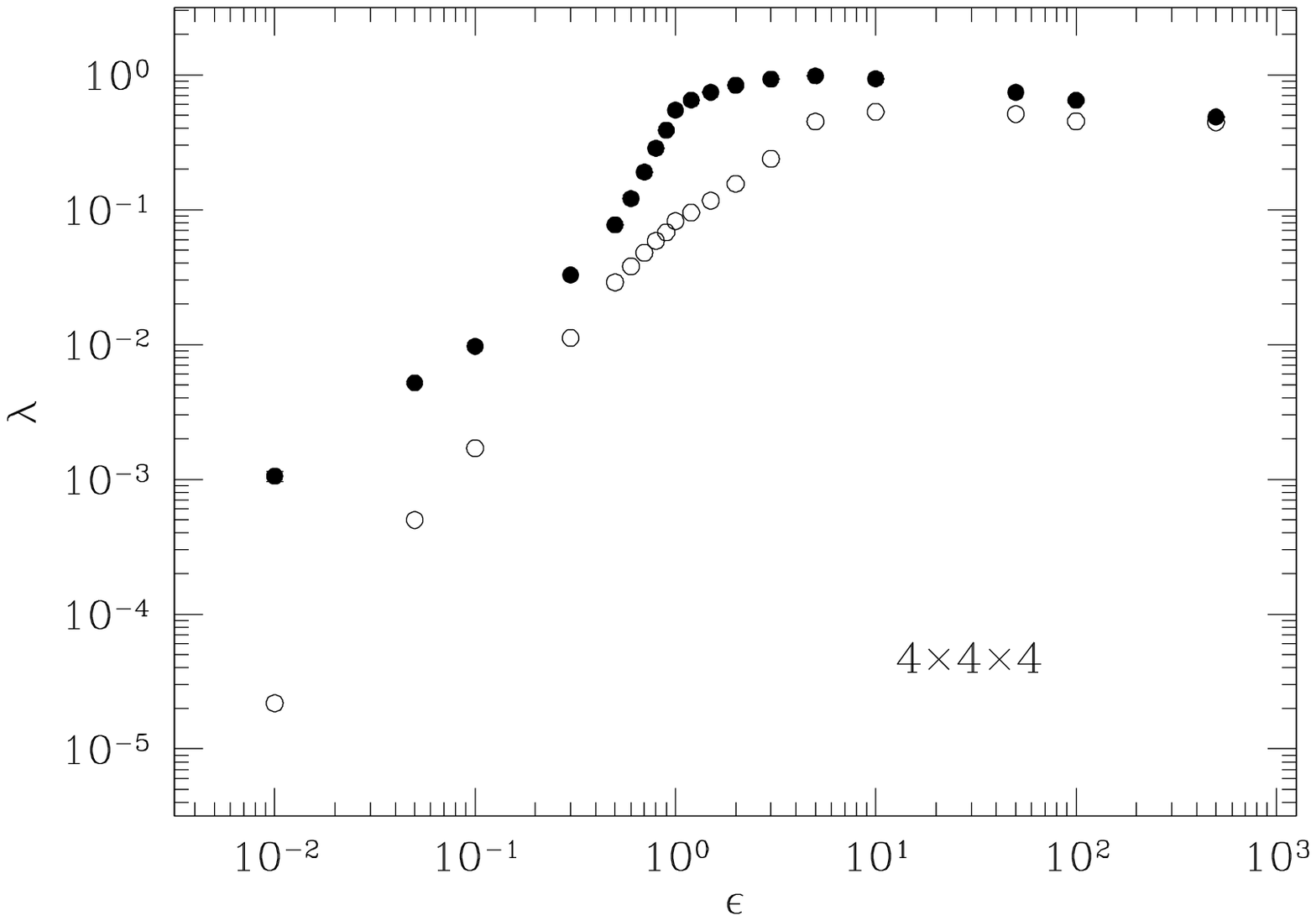,height=6.5cm}}
\centerline{\psfig{file=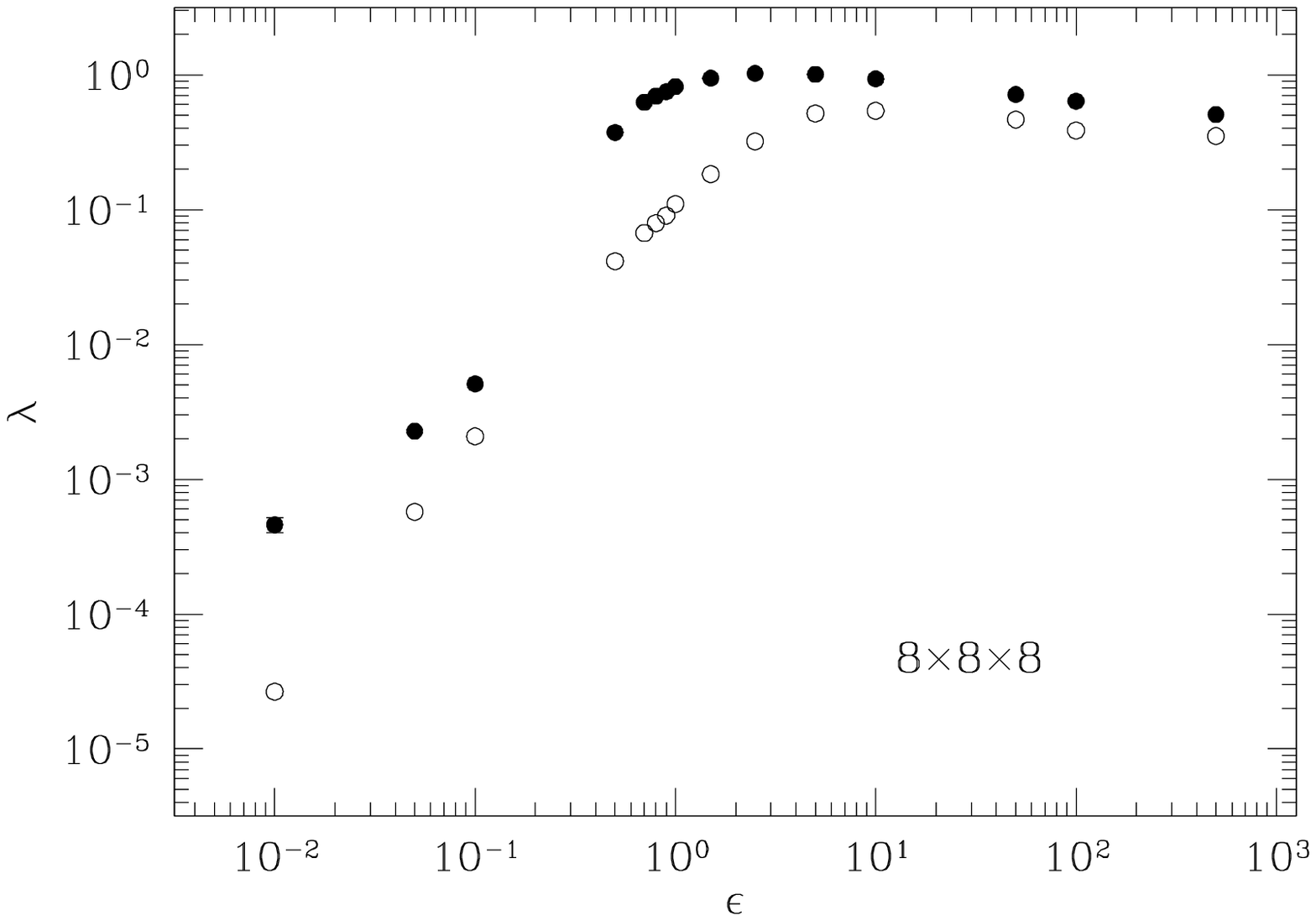,height=6.5cm}}
\centerline{\psfig{file=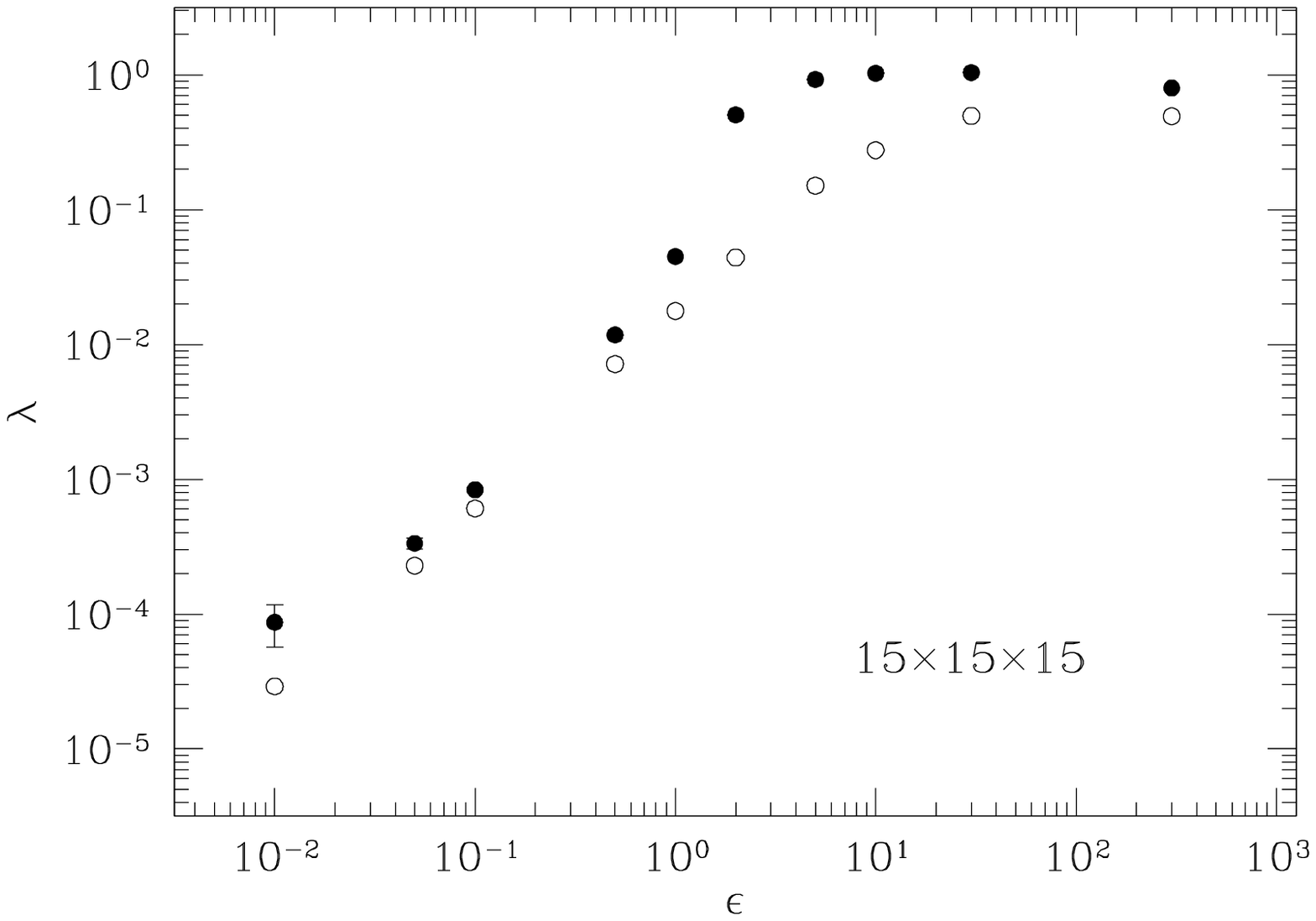,height=6.5cm}}
\caption{Largest Lyapunov exponent {\em vs.}
the energy density $\varepsilon$
for three different lattice sizes: from top to bottom,
$L^3 = 4^3,8^3,15^3$, in a log-log scale.
The full circles are the numerically measured
values, and the open circles are the theoretical estimates obtained
by inserting the average and fluctuations of the Ricci curvature, reported
in Fig. \protect\ref{fig_k_sigma}, in Eq. (\protect\ref{formula}).
The low-energy behaviour $\lambda \propto \varepsilon^2$ is well evident
from the theoretical estimates already for a $4^3$ lattice, where the scaling
with $\varepsilon$ of the numerically computed Lyapunov exponents is completely
different.
\label{fig_lyap_log}}
\end{figure}

From the arguments reported at the end of Sec. \ref{sec_geom} we expect
a crossover from weak to strong chaos where
$\sigma_k \simeq k$. The results reported in Fig. \ref{fig_k_sigma} thus
suggest that such a crossover should occur around $\varepsilon \simeq 1$,
perhaps at
a somewhat higher value of $\varepsilon$ in the case of a $15^3$ lattice.
Actually, as it is shown in Fig. \ref{fig_lyap},
in the energy region around
$\varepsilon \simeq 1$ the numerically computed values of the largest Lyapunov
exponent rapidly grow. At larger energies the values of $\lambda$ decrease
because as $\varepsilon \to \infty$ the model becomes integrable.
At variance with what is usually found in other models (where the
$\lambda(\varepsilon)$ pattern is rather stable when the number of degrees
of freedom of the system is varied) sizeable variations in the $\lambda
(\varepsilon)$ pattern are observed here at different lattice sizes.
Again, the phenomenology is very close to the one observed in the XY
model on a $2-d$ lattice. The fact that there is a change in the chaotic
properties of the dynamics is well evident
looking at Fig. \ref{fig_lyap_log}, where the Lyapunov exponents are plotted
against $\varepsilon$ in a log-log scale. In fact, in an energy range around
$\varepsilon \simeq 1$, the behaviour of $\lambda$ as a function of the energy
density deviates from a steep power law to a smoother one. In Fig.
\ref{fig_lyap_log} the numerical values of $\lambda$ are compared with the
theoretical estimates obtained according to the geometric theory outlined
in Sec. \ref{sec_geom}, i.e., obtained by substituting the computed values of
$k$ and $\sigma_k$, shown in Fig. \ref{fig_k_sigma}, into Eq. (\ref{formula}).

Two facts are immediately evident from Fig. \ref{fig_lyap_log}. First,
in the low-energy region (weak chaos)
the theoretical estimates show a power-law behaviour $\lambda \propto
\varepsilon^2$ already using the geometric values computed using a very small
($4^3$) lattice, and the exponent of the power law remains
the same for all the lattice sizes, whereas the numerical values of $\lambda$
show a less steep $\varepsilon$-dependence for small lattices.
Second, both the low-energy power law behaviour and the actual values of
the numerical Lyapunov exponents are the closer to the theoretical
estimates the
larger the lattice size is.
Let us mention that the actual value of $\lambda$ is theoretically
underestimated in the transition region, as it is well evident from
Fig. \ref{fig_lyap_log}. Without entering the details of this mismatch,
we mention that the reason can be understood in the light of a very similar
situation found in $1-d$, $2-d$ and $3-d$ XY models \cite{pre96,cccp,gatto}.

\begin{figure}
\centerline{\psfig{file=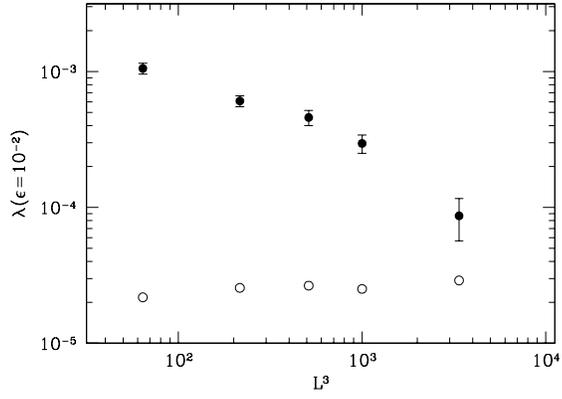,height=6.5cm}}
\caption{Numerical results for the Lyapunov exponent at $\varepsilon = 10^{-2}$
(full circles) compared with the theoretical estimate obtained by
Eq. (\protect\ref{formula}) (open circles) {\em vs.} lattice size.
\label{fig_lyap_N}}
\end{figure}

Moreover, the actual values of the theoretical
estimates are almost free from finite-size effects, which are instead very
large in the numerical values of the Lyapunov exponents.
This fact is particulalrly evident at very low energy densities, as shown
in Fig. \ref{fig_lyap_N}, where the numerical and theoretical values
of $\lambda$ are plotted against the lattice size.

\section{Concluding remarks}
\label{sec_remarks}

The results presented in the previous section clearly
show that the geometric approach to
Hamiltonian chaotic dynamics outlined in Sec.\ref{sec_geom} is particularly
efficient to provide very good estimates of the largest Lyapunov exponent
for a classical lattice gauge theory with abelian gauge symmetry.
In particular, the theoretical estimates of the Lyapunov exponents are
almost free from finite-size effects, especially at low energies.
The theoretical values are
very close to the numerical
values that one extrapolates for very large lattice sizes
even when the geometric quantities are computed using very small lattices.
Also the dependence of $\lambda$ on the energy density in the weakly chaotic
(low-energy) region is $\lambda \propto \varepsilon^2$ already for
a $4^3$ lattice, where the numerical values of the $\lambda$'s exhibit a
much less steep scaling.

This is a confirmation of the fact that the geometric estimates of
the Lyapunov exponents obtained by Eq. (\ref{formula}) are very
good for large systems, as already observed in other cases \cite{pre96}.
It is not surprising since the whole theory uses the large size of
the systems as a hypothesis \cite{pre96,physrep}. Moreover, this feature
is very promising in the perspective of applying the geometric
theory to non-abelian lattice gauge theories. In fact, as already
mentioned in the Introduction, in the non-abelian case there is still some
doubt about the correct value of the exponent $\alpha$ of the power law
$\lambda \propto \varepsilon^\alpha$ in the low-energy region,
and this has remarkable importance for the relevance of chaos in the
continuum limit of the theory. The application of the geometric theory of chaos
might perhaps help in obtaining reliable estimates of the exponent $\alpha$
already using small lattice sizes.

\acknowledgments

We thank G. Pettini for
fruitful discussions. LC
acknowledges also useful discussions with A. Giansanti.
The present work was mainly carried out during one of the authors' (LC)
stay at the D\'epartement de Physique Th\'eorique de l'Universit\'e de
Gen\`eve which is gratefully acknowledged for the kind hospitality. This work
is part of the EEC project CHRXCT94/0579 (OFES 95.0200).

\newpage

\end{document}